\documentclass[journal]{IEEEtran}
\usepackage{graphicx}
\usepackage{amsfonts,amssymb,amsmath,mleftright}
\usepackage{xcolor}
\usepackage{bm,bbm}
\usepackage{booktabs}
\usepackage{multirow}
\usepackage{siunitx}
\usepackage{float}
\usepackage{subfig}
\usepackage{mathtools}
\usepackage{adjustbox}
\graphicspath{{images/}}

\usepackage{algorithm}
\usepackage{algpseudocode}

\ifCLASSINFOpdf
\else
\fi
\begin{document}

\newcommand{\red}{\textcolor{red}}
\newcommand{\blue}{\textcolor{black}}

%
\title{Estimation of Fish Catch Using Sentinel-2, 3 and XGBoost-Kernel-Based Kernel Ridge Regression}
%
%
%

\author{Kanu~Mohammed,
    Vaishnavi~Joshi,
    Pranjali~Diliprao~Patil,
    Sandipan~Mondal,
    Ming-An~Lee$^{*}$,
	Subhadip~Dey$^{*}$,~\IEEEmembership{Senior~Member,~IEEE}
	\thanks{K.~Mohammed, V.~Joshi, P.~D.~Patil and S.~Dey are with the Optical and SAR Sensing Laboratory, Agricultural and Food Engineering Department, Indian Institute of Technology Kharagpur, Kharagpur, India, (e-mail(s): kanumohamed0@gmail.com, vaishnavijoshi2701@gmail.com, sdey23@agfe.iitkgp.ac.in).}
    \thanks{S.~Mondal and M.-A.~Lee are with Department of Environmental Biology and Fishery Science, National Taiwan Ocean University, Keelung City, Taiwan, Center of Excellence for the Oceans, National Taiwan Ocean University, Keelung City, Taiwan, (e-mail: mondalsandipan31@gmail.com, malee@mail.ntou.edu.tw).}
    \thanks{$^*$Corresponding authors}
}

\maketitle

\begin{abstract}
Oceanographic factors, such as sea surface temperature and upper-ocean dynamics, have a significant impact on fish distribution. Maintaining fisheries that contribute to global food security requires quantifying these connections. This study uses multispectral images from Sentinel-2 MSI and Sentinel-3 OLCI to estimate fish catch using an Extreme Gradient Boosting (XGBoost)–kernelized Kernel Ridge Regression (KRR) technique. According to model evaluation, the XGBoost-KRR framework gets the strongest correlation and the lowest prediction error across both sensors, suggesting improved capacity to capture nonlinear ocean–fish connections. While Sentinel-2 MSI resolves finer-scale spatial variability, emphasizing localized ecological interactions, Sentinel-3 OLCI displays smoother spectral responses associated with poorer spatial resolution. By supporting sustainable ecosystem management and bolstering satellite-based fisheries assessment, the suggested approach advances SDGs 2 (Zero Hunger) and 14 (Life Below Water).   
\end{abstract}


%
\IEEEpeerreviewmaketitle

\section{Introduction}
\label{Sec:Introduction}
Aquaculture plays an important role in addressing global issues of food and nutritional security,  for which aquaculture and capture fisheries are often studied as parallel domains, yet their interlinkages are increasingly significant in ensuring sustainable fish production. However, understanding the correlation between the fish and the environment is not simple. In literature, the fish distribution is often related with the Sea Surface Temperature (SST) and ocean states~\cite{sun2020retrieval, liu2018assimilating}. Williams~\cite{williams1977sea} showed that the tuna fisheries of Australia is strongly related to the SST map derived from the airborne radiometry data. It was also observed that the higher concentration of fish exists mostly around the strong temperature gradient.

Favorable environmental conditions for feeding and breeding such as ocean temperature gradients, chlorophyll concentration, dissolved oxygen and depth of mixed layer are the key factors that influence the movement of fishes in the ocean~\cite{pranovi2004multidisciplinary}. Importantly, the upwelling characteristics of a marine region puts an impact on the seasonal fish concentration as the upwelling helps in producing phytoplankton which acts as a nutrient to fishes.~Lasker et al.~\cite{lasker1981use} used the infrared imagery of AVHRR to understand the relation between ocean processes and spawning of the Northern Anchovy. 

Several studies tried to find a joint correlation of SST and chlorophyll in fish production~\cite{fiedler1984fisheries, wall2008satellite}.~Solanki et al.~\cite{solanki2001application} studied the SST and chlorophyll distribution over coastal area using IRS P4 OCM and synchronous NOAA AVHRR. Further, different oceanic features like, coastal fronts, fringe fronts, rings, mushrooms and others were considered for experimental fishery forecast. In a similar direction, Dwivedi et al.~\cite{dwivedi2005exploration} explored the color gradient-based analysis to capture the fishing grounds. SST countours were overlaid onto the chlorophyll image to capture the common frontal structure. Fisheries forecast were generated within 24 hours of the satellite overpass.

The habitat stability models, especially the species distribution model were often used to understand the spatio-temporal relationship between species and habitat which can be useful to infer the liklihood of occurence of species abundance for a given habitat attributes~\cite{guisan2005predicting}. These semi-empirical models also help to assess the distribution patterns at different scales. For example, gradual distributions are observed over a large extent and mostly controlled by the climatic factors, while, over a small area, the governing factors are mostly the micro-topographic variations or, habitat fragmentation~\cite{raven2002predicting}. In this regard, Vincenzi et al.~\cite{vincenzi2006estimating, vincenzi2007comparative} used semi-empirical and zero-inflated regression models to predict the yield potential of \textit{R. philippinarum}. Later, the same authors~\cite{vincenzi2011application} used the random forest algorithm to predict the potential spatial yield using the environmental factors, such as, percentage of sand in the sediment, salinity, and water depth.

In recent years, Azeez et al.~\cite{abdul2021predicting} used Generalized Additive Model (GAM) and Boosted Regression Tree (BRT) to determine the potential fishing ground of ribbon fish (\textit{Trichiurus lepturus}). In their study, MODIS Aqua data were used along with the euphotic depth, SST, bathymetry and sea surface height anomaly. Among them, SST and bathymetry were found to be the two most important parameters to understand the fish distribution. Later, Purwanto et al.~\cite{purwanto2024seasonal} used the remote sensing data from Suomi-National Polar-Orbiting Partnership-Visible Infrared Imaging Radiometer Suite to determine the potential fishing zone around the coast of Indonesia. Single Image Edge Detection was applied to detect the zones and it was found that these zones had a strong correlation with the upwelling pattern.

Therefore, it can be seen that the existing studies mostly performs either simple regression techniques or, tree-based techniques. The primary limitation of only tree-based regression technique is the piecewise constant predictions and the bias is implicit and heuristic. Hence, in this study, we use the Kernel Ridge Regression (KRR) technique along with the trained XGBoost-based kernel to estimate the fish catch from the Sentinel-2 dataset.  

\section{Methodology}
\label{Sec:Methodology}

In this manuscript, we use an Extreme Gradient Boosting (XGBoost) based Kernel Ridge Regression (KRR) method to estimate fish catch using Sentinel-2 dataset. KRR is a non-linear ridge regression technique where kernel is used to capture the complex pattern of the dataset. A standard linear ridge regression tries to map the covariates $\mathbf{x}_i \in \mathbb{R}^{\text{M}}$ to the response variable $y_i\in\mathbb{R}$ as, $\phi: \mathbb{R}^{\text{M}} \rightarrow \mathcal{H}$, where, $\mathcal{H}$ is the Hilbert space and $i \in [1, \text{N}]$. A linear model in the feature space can be written as, $f(\mathbf{x}) = \mathbf{w}^{\text{T}}\phi(\mathbf{x})$, where, $\mathbf{w}$ is the weight. In this regard, the classical solution tries to minimize the squared loss between the response variable and the predicted value, i.e., $\mathbf{\Phi}\mathbf{w}$.    

\begin{equation}
   \min_{\mathbf{w}} \hspace{2mm} \lVert\mathbf{y} - \mathbf{\Phi}\mathbf{w}\rVert^2_2
\end{equation}

where, $\mathbf{\Phi} = [\phi(\mathbf{x}_1)^\text{T} \hspace{2mm} \phi(\mathbf{x}_2)^\text{T} \hspace{2mm} \cdots \hspace{2mm} \phi(\mathbf{x}_N)^\text{T}]^\text{T}$. 
However, due to the scarcity of the training samples, the estimated variance of $\mathbf{w}$ might be large and hence, the reliability of the estimated values may be questionable. One way to minimize the variance of $\mathbf{w}$ is to penalize the second-order norm of $\mathbf{w}$. Hence, the ridge regression uses the following cost function:

\begin{equation}
    \min_{\mathbf{w}} \hspace{2mm} \lVert\mathbf{y} - \mathbf{\Phi}\mathbf{w}\rVert^2_2 + \lambda\lVert \mathbf{w}\rVert^2_2
\end{equation}

The hyper-parameter $\lambda > 0$ allows higher training error to achieve a more reliable variance estimate. It also controls the trade-off between the bias and variance of the estimates. In general, cross-validation technique can be applied to obtain the optimum value of the regularization parameter $\lambda$. 

The optimal solution satisfies,
\begin{equation}
    \mathbf{w} = \mathbf{\Phi}^\text{T}\mathbf{\alpha}
\end{equation}

where, $\mathbf{\alpha} \in \mathbb{R}^\text{N}$ is known as kernel weights. Therefore, the predicted response values can be written as,
\begin{equation}
    \hat{\mathbf{y}} = \mathbf{\Phi}\mathbf{w} = \mathbf{\Phi}\mathbf{\Phi}^\text{T}\mathbf{\alpha}
\end{equation}

Here, $\mathbf{\Phi}\mathbf{\Phi}^\text{T}$ can be represented as a kernel, $\mathbf{K} = \mathbf{\Phi}\mathbf{\Phi}^\text{T} \in \mathbb{R}^{\text{N} \times \text{N}}$. As a result, the cost function can be rewritten as,
\begin{equation}
    \min_{\mathbf{\alpha}} \hspace{2mm} \lVert\mathbf{y} - \mathbf{K}\mathbf{\alpha}\rVert^2_2 + \lambda\mathbf{\alpha}^\text{T}\mathbf{K}\mathbf{\alpha}
\end{equation}

After taking the derivative with respect to $\mathbf{\alpha}$ in the closed-form solution we get,
\begin{equation}
    \mathbf{\alpha} = (\mathbf{K} + \lambda\mathbf{I})^\text{T}\mathbf{y} \in \mathbb{R}^{\text{N}}
\end{equation}

During prediction if the test data $\mathbf{x}_*$ is of $\mathbb{R}^\text{M}$, the kernel can be written as, $\mathbf{k}_* = [k(\mathbf{x}_1, \mathbf{x}_*) \hspace{2mm} \cdots \hspace{2mm} k(\mathbf{x}_\text{N}, \mathbf{x}_*)]^\text{T}$, where, $\mathbf{k}_* \in \mathbb{R}^\text{N}$. The predicted response can be written as, 
\begin{equation}
    \hat{\mathbf{y}}_* = \mathbf{k}_*^\text{T}\mathbf{\alpha}
\end{equation}

In general, the kernel can be linear or, radial basis function type. In this manuscript, we define the kernel from the trained XGBoost model of $\text{P}$ trees. Let's assume that p-th tree has $L_p$ leaves. Then the leaf index function of the p-th tree can be written as, $f_p(\mathbf{x_i}) \in \{1, 2, \cdots, L_p\}$. We define the one-hot embedding for the p-th tree as,
\begin{equation}
    z^{(p)}_{l}(\mathbf{x}_i) =
\begin{cases}
1, & \text{if } f_p(\mathbf{x}_i) = l, \\
0, & \text{otherwise}.
\end{cases}
\end{equation}

where, $l = \{1, 2, \cdots , L_p\}$. The concatenated tree embeddings can be written as,

\begin{equation}
    z(\mathbf{x}_i) = [z^{(1)}(\mathbf{x}_i), \hspace{2mm} \cdots, \hspace{2mm} z^{(P)}(\mathbf{x}_i)]^{\text{T}}
\end{equation}

where, $z(\mathbf{x}_i) \in \mathbb{R}^\text{L}$, $\text{L} = \sum_{p = 1}^P L_p$. Therefore, the complete embedding matrix can be expressed as,
\begin{equation}
    \mathbf{Z} = \begin{bmatrix}
        z(\mathbf{x}_1)^\text{T}\\
        z(\mathbf{x}_2)^\text{T}\\
        \vdots\\
        z(\mathbf{x}_N)^\text{T}
    \end{bmatrix}
\end{equation}

where, $\mathbf{Z} \in \mathbb{R}^{\text{N} \times \text{L}}$. Further, the kernel, $\mathbf{K}$ can be defined as, $\mathbf{K} = \dfrac{1}{P}\mathbf{Z}\mathbf{Z}^{\text{T}}$.

%
%
%
%
%
%
\section{Dataset}
\label{Sec:dataset}

\begin{figure*}[hbt]
\centering
\subfloat[Sentinel- 2\label{fig:violin_plot_spectral_bands_S2}]{%
	\includegraphics[width=\textwidth]{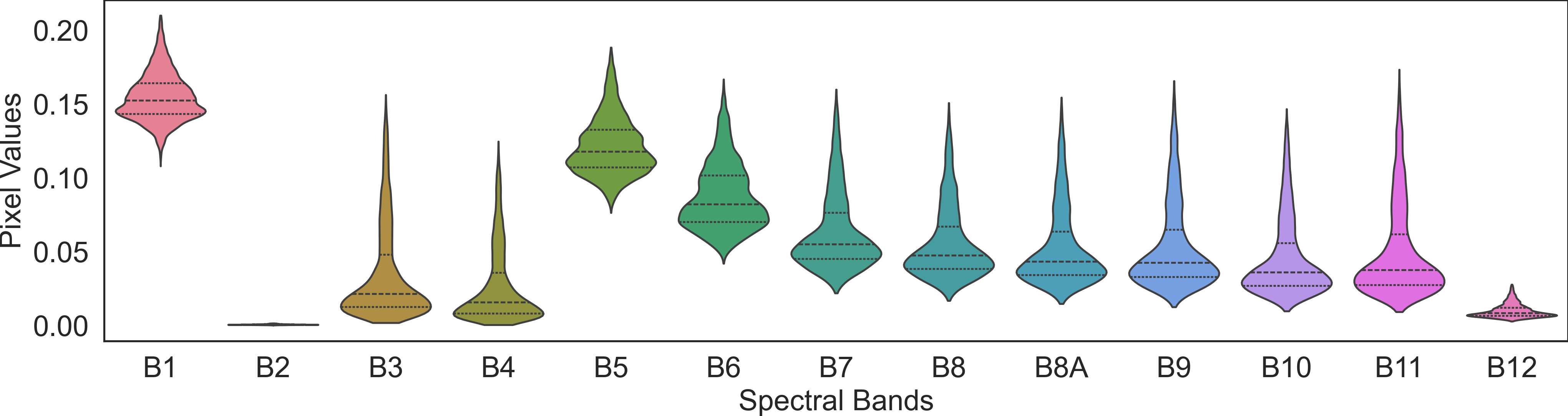}}
\hfill
\subfloat[Sentinel - 3\label{fig:S3_violinPlot}]{%
	\includegraphics[width=\textwidth]{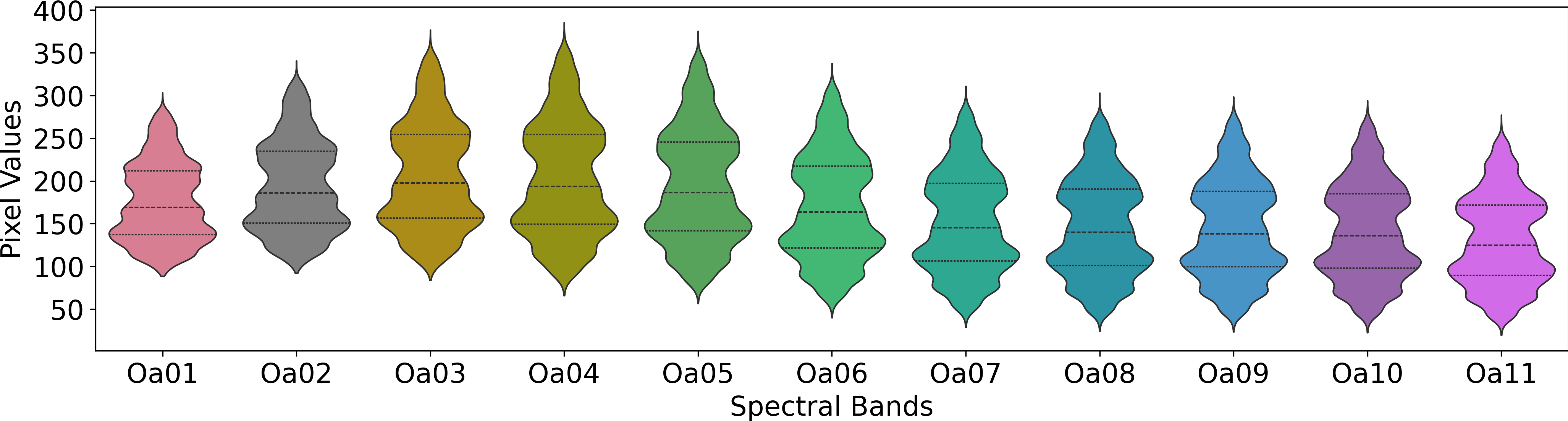}}
\caption{Violin plots of different spectral bands of (a) Sentinel-2 and, (b) Sentinel-3 data. The pixel values of Sentinel-2 are reflectance and of Sentinel-3 are radiance values.}
\label{fig: violin_plots}
\end{figure*}



\begin{figure}[!h]
\centering
  \includegraphics[width=\columnwidth]{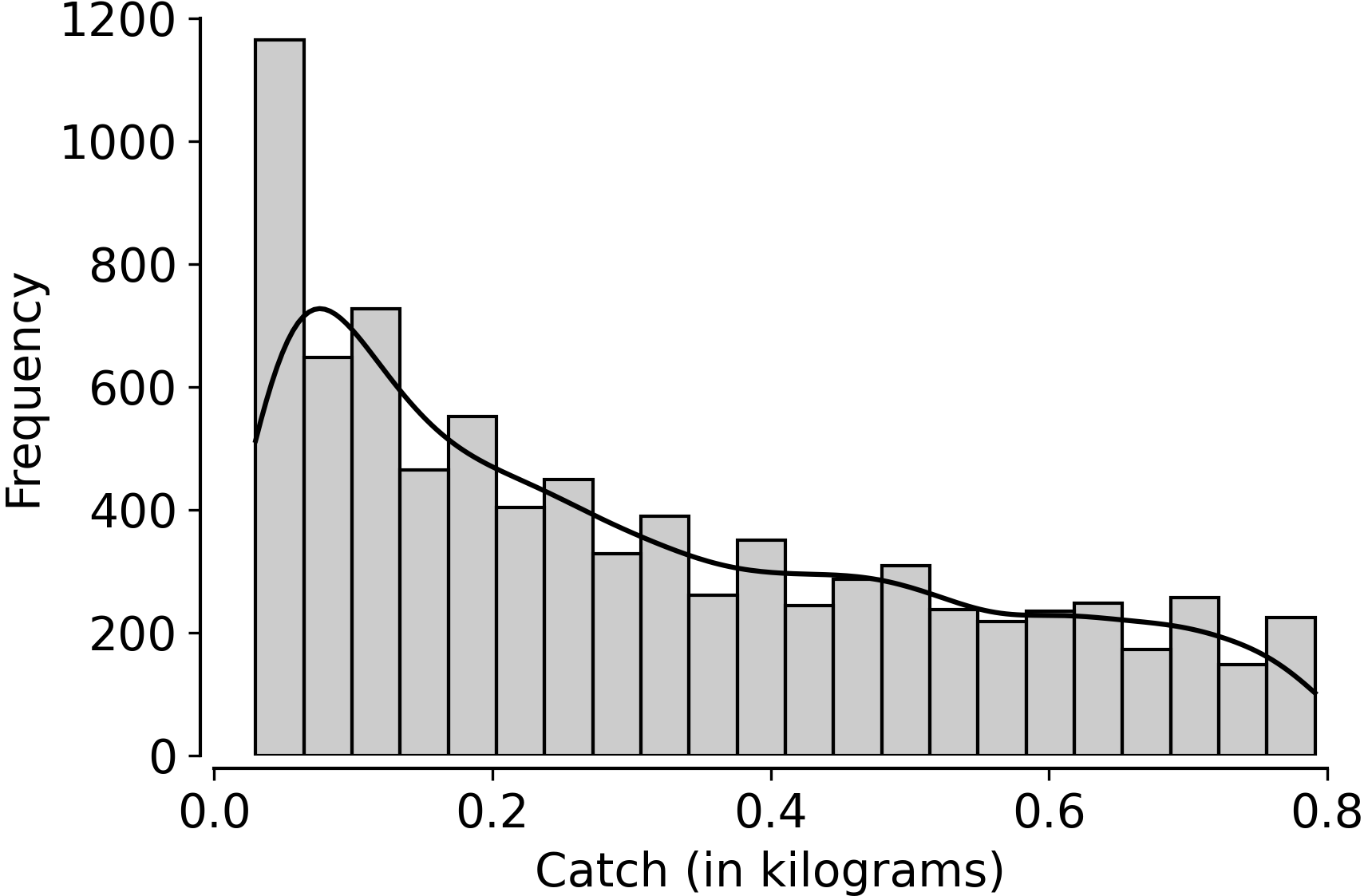}
  \caption{The distribution of fish catch dataset in kilograms used in this study. The thick dark line is the kernel density estimation of the data}
  \label{fig:fish_catch_dist}
\end{figure}

The monthly logbook data for the Taiwanese fishery targeting seabreams in the Taiwan coastal waters, namely within the geographical area bounded by \ang{22} N, \ang{117} E and \ang{27} N, \ang{126} E, are obtained from the Fisheries Agency, Ministry of Agriculture, Taiwan. The research is conducted until December 2019 with \ang{0.1} spatial resolution. The dataset includes information on the fishing date (year and month), fishing location (latitude and longitude), catch (in kilograms), fishing gear used, and vessel tonnage. This research examined the catch data that has the most substantial impact in capturing scenario. Fig.~\ref{fig:fish_catch_dist} shows the distribution of \textit{in-situ} fish catch data in kilograms. It can be seen that the catch varies from $\approx$ \SIrange[]{0}{0.8}{\kilo\gram} with a positive skewness. The decreasing heights of bar across the range depicts a exponential trend where biomass distribution declines with size. Moreover, it shows a size-abundance relationship which are common in typical ecosystems. Based on the availability of Sentinel-2 and Sentinel-3 data and cloud cover over pixels, the total number of dataset for Sentinel-2 and Sentinel-3 differs. Finally, we clip the histograms of fish catch data for individual Sentinel-2 and Sentinel-3 acquisitions within \SI{10}{\percent} to \SI{90}{\percent}. These dataset are used for the regression analysis using the Kernel Ridge Regression techniques.

\section{Results and Discussion}
\label{Sec:Result}

Figure~\ref{fig: violin_plots} presents symmetric violin plots of pixel value distributions for Sentinel-2 (top panel) and Sentinel-3 OLCI (bottom panel) spectral bands extracted over the ocean surface within the study area, corresponding to fish catch values between 0.2 and 0.8~kg. Each violin illustrates the density, spread, and central tendency of individual spectral bands, thereby highlighting wavelength-dependent sensitivity to ocean-surface and upper-water-column conditions associated with fish catch variability.

For Sentinel-2, the coastal aerosol band (B1, 443~nm) shows comparatively higher reflectance values with a relatively narrow distribution. This behaviour reflects its dominant sensitivity to atmospheric scattering and surface reflectance rather than to subsurface oceanic constituents. The blue band (B2, 490~nm) exhibits low pixel values and minimal spread, consistent with strong absorption by seawater and chlorophyll and limited penetration depth. The green band (B3, 560~nm) shows a broader distribution than B2, indicating greater sensitivity to variations in suspended matter and near-surface turbidity. This behaviour is expected, as green wavelengths penetrate the ocean surface more effectively and provide improved contrast under varying optical conditions. The red band (B4, 665~nm) exhibits low to moderate reflectance with limited variability, reflecting strong chlorophyll absorption and reduced water penetration in the red region. A clear increase in variability is observed in the red-edge bands (B5-B7; 705-783~nm), which exhibit wider violin shapes and higher median values compared to the visible bands. These wavelengths are sensitive to chlorophyll concentration and surface phytoplankton dynamics, and the observed spread indicates spatial heterogeneity in biological productivity across the ocean surface. The near-infrared band (B8, 842~nm) and narrow near-infrared band (B8A, 865~nm) show moderate reflectance values with relatively stable distributions. As seawater strongly absorbs near-infrared radiation, the observed variability is primarily influenced by surface scattering effects related to phytoplankton presence, organic matter, and surface roughness rather than subsurface properties. The water vapour band (B9, 945~nm) and cirrus band (B10, 1375~nm) exhibit low reflectance and narrow distributions, indicating limited sensitivity to ocean-surface conditions. Similarly, the shortwave infrared bands (B11 at 1610~nm and B12 at 2190~nm) show very low pixel values and compact distributions due to strong water absorption, restricting their relevance for direct ocean-water characterization.

In the Sentinel-3 OLCI data, all bands exhibit broader and smoother violin shapes compared to Sentinel-2, reflecting the coarser spatial resolution of 300~m and the spatial averaging of ocean-surface features within individual pixels. The short-wavelength blue bands (Oa01-Oa03; 400-442.5~nm) show elevated pixel values and wide distributions, indicating sensitivity to aerosols, coloured dissolved organic matter, and chlorophyll absorption features. The smooth and symmetric shapes suggest stable radiometric behaviour across the study area, with variability driven by differences in oceanic water composition rather than extreme outliers. The blue-green bands (Oa04-Oa06; 490-560~nm) display the widest distributions among the OLCI bands, highlighting their strong sensitivity to chlorophyll concentration, suspended particles, and turbidity in surface waters. These bands show a clear spread in pixel values, indicating spatial heterogeneity in nutrient availability and biological activity across the ocean surface. Their broader distributions, relative to the blue bands, indicate increased responsiveness to in-water optical properties closely linked to marine productivity. The red and red-edge bands (Oa07-Oa11; 620-708.75~nm) exhibit comparatively narrower but structured distributions, with visible density peaks indicating stable but varying surface biological conditions. These wavelengths correspond to chlorophyll absorption maxima and fluorescence-related regions, and the observed variability reflects differences in phytoplankton concentration and surface biomass. Despite increasing water absorption toward longer wavelengths, the persistence of measurable spread in these bands indicates that OLCI retains sensitivity to surface biological signals even at coarser spatial resolution.

Overall, the Sentinel-3 violin plots demonstrate strong spectral sensitivity to bulk water-quality parameters governing productivity within the study area, while the smoothness of the distributions reflects spatial averaging effects. This behaviour contrasts with Sentinel-2, which captures finer-scale spatial variability, and reinforces the role of Sentinel-3 for characterising surface and water-column conditions associated with fish catch variability. In this letter, we have used bands 2, 3, 4, 5, 6, 7, 8, 8A, 9 for Sentinel-2 data and bands 3, 4, 7, 8, 9, 10 for Sentinel-3 data as they possess unique characteristics depending on the oceanic biological conditions. We compared the results using three different kernels in Kernel Ridge Regression (KRR), i.e., Linear Kernel (KRR-Lin), Radial Basis Function Kernel (KRR-RBF), and, Extreme Gradient Boosting Kernel (KRR-XGB).

\begin{table}[hbt]
\centering
\caption{The Root Mean Square Error (RMSE), Correlation coefficient ($\rho$), p-value and D-value of three different kernels with KRR for Sentinel-2 data}
\label{tab:stat_S2}
\begin{tabular}{ccccc}
\toprule
\multicolumn{1}{l}{\textbf{\begin{tabular}[c]{@{}l@{}}Algorithm \end{tabular}}} & \multicolumn{1}{c}{\textbf{RMSE}} & \multicolumn{1}{c}{\textbf{$\rho$}} & \multicolumn{1}{c}{\textbf{p-value}} & \multicolumn{1}{c}{\textbf{D-value}} \\
\midrule
KRR-Lin & 0.218 & -0.032 & 2.48 $\times 10^{-4}$ & 0.275 \\
KRR-RBF & 0.210 & 0.069 & 1.26 $\times 10^{-3}$ & 0.239 \\
KRR-XGB & \textbf{0.085} & \textbf{0.924} & 2.09 $\times 10^{-61}$ & \textbf{0.952} \\
\bottomrule
\end{tabular}
\end{table}

\begin{table}[hbt]
\centering
\caption{The Root Mean Square Error (RMSE), Correlation coefficient ($\rho$), p-value and D-value of three different kernels with KRR for Sentinell-3 data}
\label{tab:stat_S3}
\begin{tabular}{ccccc}
\toprule
\multicolumn{1}{l}{\textbf{\begin{tabular}[c]{@{}l@{}}Algorithm \end{tabular}}} & \multicolumn{1}{c}{\textbf{RMSE}} & \multicolumn{1}{c}{\textbf{$\rho$}} & \multicolumn{1}{c}{\textbf{p-value}} & \multicolumn{1}{c}{\textbf{D-value}} \\
\midrule
KRR-Lin & 0.194 & 0.023 & 5.19 $\times 10^{-6}$ & 0.406 \\
KRR-RBF & 0.160 & 0.021 & 5.6 $\times 10^{-2}$ & 0.448 \\
KRR-XGB & \textbf{0.116} & \textbf{0.731} & 2.98 $\times 10^{-60}$ & \textbf{0.771} \\
\bottomrule
\end{tabular}
\end{table}

The statistical comparisons of KRR-Lin, KRR-RBF and KRR-XGB are shown in Table~\ref{tab:stat_S2} for Sentinel-2 data and in Table~\ref{tab:stat_S3} for Sentinel-3 data. For Sentinel-2 data the RMSE values for KRR-Lin and KRR-RBF are \SI{0.218}{} and \SI{0.210}{}, respectively, while the RMSE for KRR-XGB is \SI{0.085}{}. Therefore, with respect to minimization of error, KRR-XGB provides \SIrange[]{59}{61}{\percent} lesser error compared to KRR-Lin and KRR-RBF. Moreover, a significant improvement in $\rho$ and D-value are observed when KRR-XGB is used instead of KRR-Lin and KRR-RBF. An enhancement of \SI{246}{\percent} and \SI{298}{\percent} in D-values is observed for KRR-XGB with respect to KRR-Lin and KRR-RBF, respectively.

For Sentinel-3 data the RMSE values for KRR-Lin and KRR-RBF are \SI{0.194}{} and \SI{0.160}{}, respectively, while the RMSE for KRR-XGB is \SI{0.116}{}. Therefore, with respect to minimization of error, KRR-XGB provides \SIrange[]{27.6}{40}{\percent} lesser error compared to KRR-Lin and KRR-RBF. Moreover, a significant improvement in $\rho$ and D-value are observed when KRR-XGB is used instead of KRR-Lin and KRR-RBF. An enhancement of \SI{89}{\percent} and \SI{72}{\percent} in D-values are observed for KRR-XGB with respect to KRR-Lin and KRR-RBF, respectively.

\begin{figure}[hbt]
\centering
\subfloat[\label{fig:correlation_plot_S2}]{%
	\includegraphics[width=0.49\columnwidth]{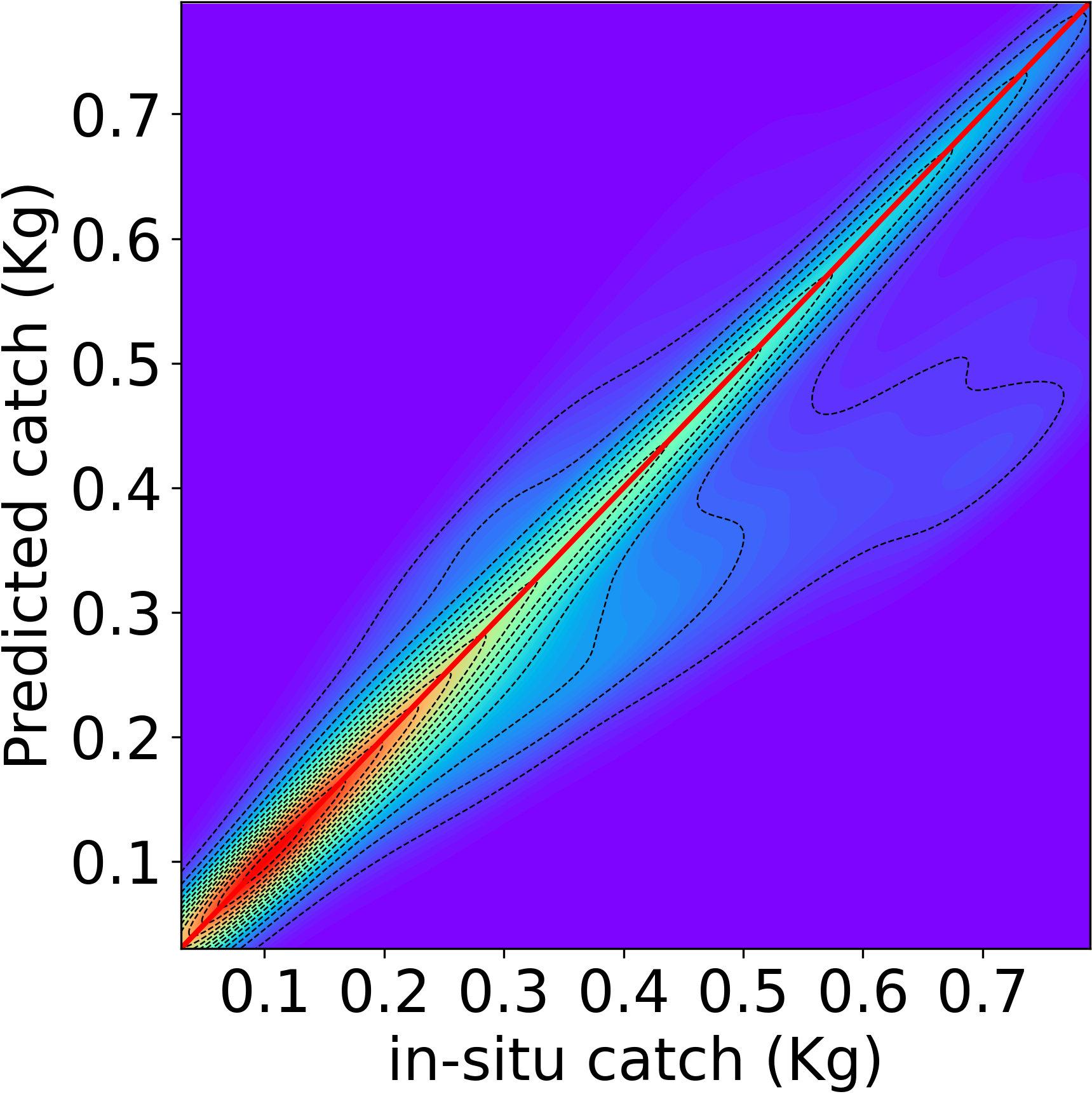}}
\hfill
\subfloat[\label{fig:correlation_plot_S3}]{%
	\includegraphics[width=0.49\columnwidth]{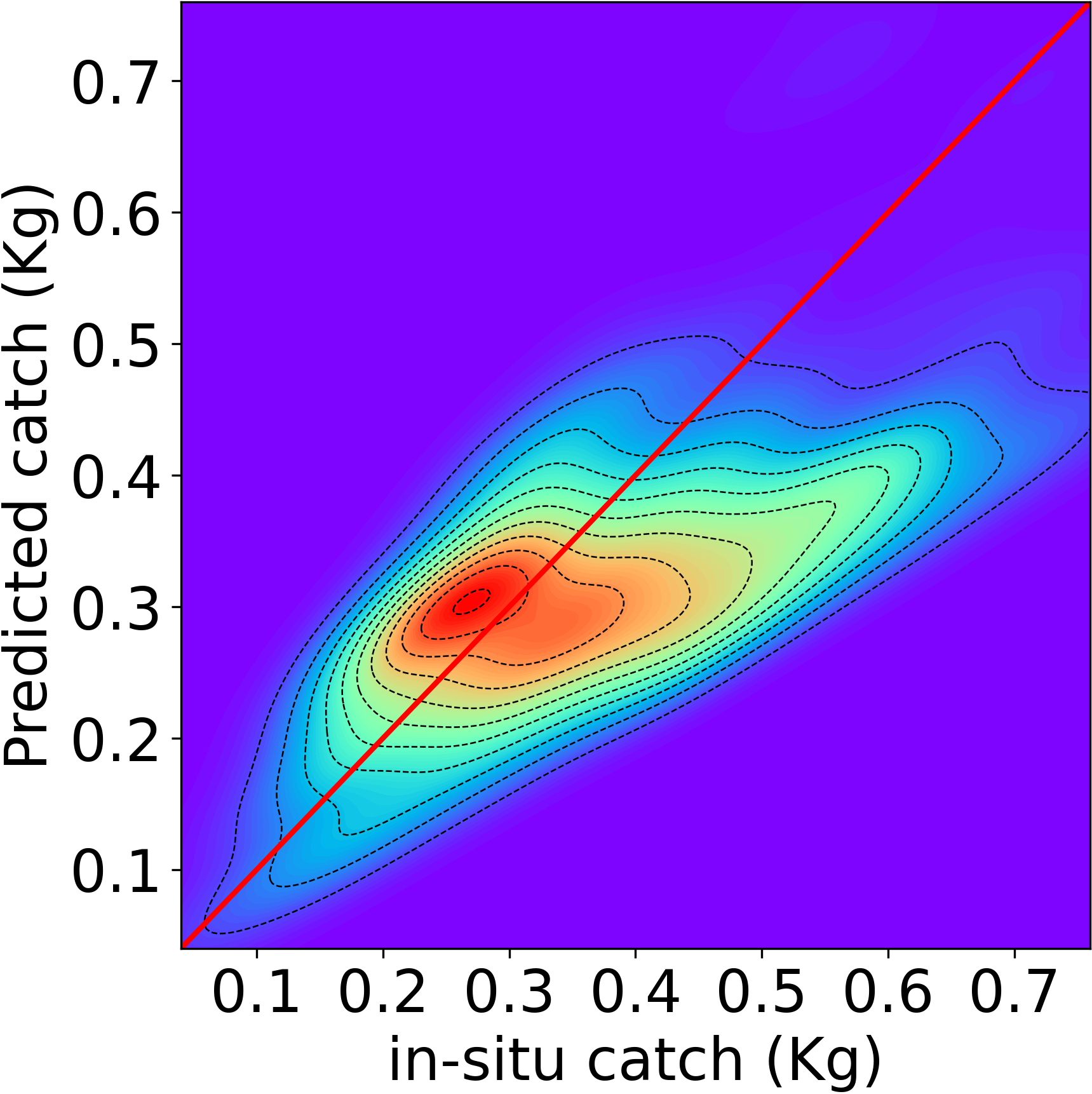}}
\caption{Correlation plots between in-situ fish catch and predicted fish catch for (a) Sentinel-2 and (b) Sentinel-3 dataset.}
\label{fig: correlation_plots}
\end{figure}

The correlation plot between in-situ fish catch and predicted fish catch is shown in Fig.~\ref{fig: correlation_plots}. For Sentinel-2 correlation plot a strong relationship exists between reference and estimated values. A high density occurs at the lower values of fish catch which indicates that the number of samples within the range is higher than the other regions and also the narrow gaps among the contours suggest low variance in that region. A slight widening of contours at mid-to-higher values indicates increasing variance and minor dispersion as magnitude increases. A small asymmetric spread on the upper side at higher values indicates a mild overestimation in that range, but no major systematic bias is evident from the correlation plot. 

The correlation plot for Sentinel-3 shows a clear positive relationship between in-situ and predicted fish catch, with most observations concentrated along the upward-sloping diagonal region. A spread around the 1:1 line indicates moderate dispersion towards the higher values. Similar to Sentinel-2 correlation plot, narrow gaps among the contours are found at lower values, suggesting low variance, while higher gaps are found in the higher values, suggesting higher variance.

\begin{figure}[hbt]
\centering
\subfloat[\label{fig:predicted_catch_S2}]{%
	\includegraphics[width=0.49\columnwidth]{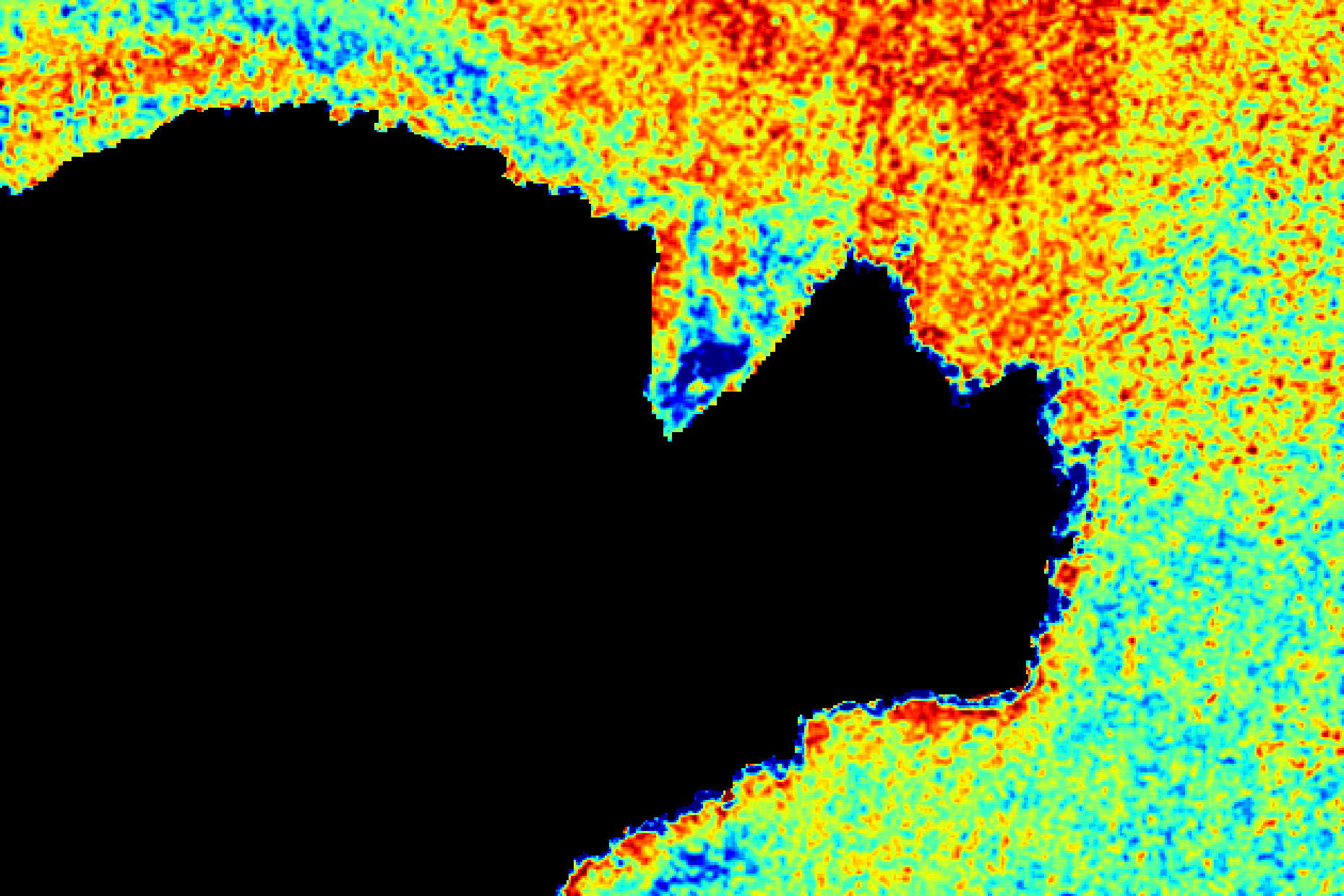}}
\hfill
\subfloat[\label{fig:predicted_catch_S3}]{%
	\includegraphics[width=0.33\columnwidth]{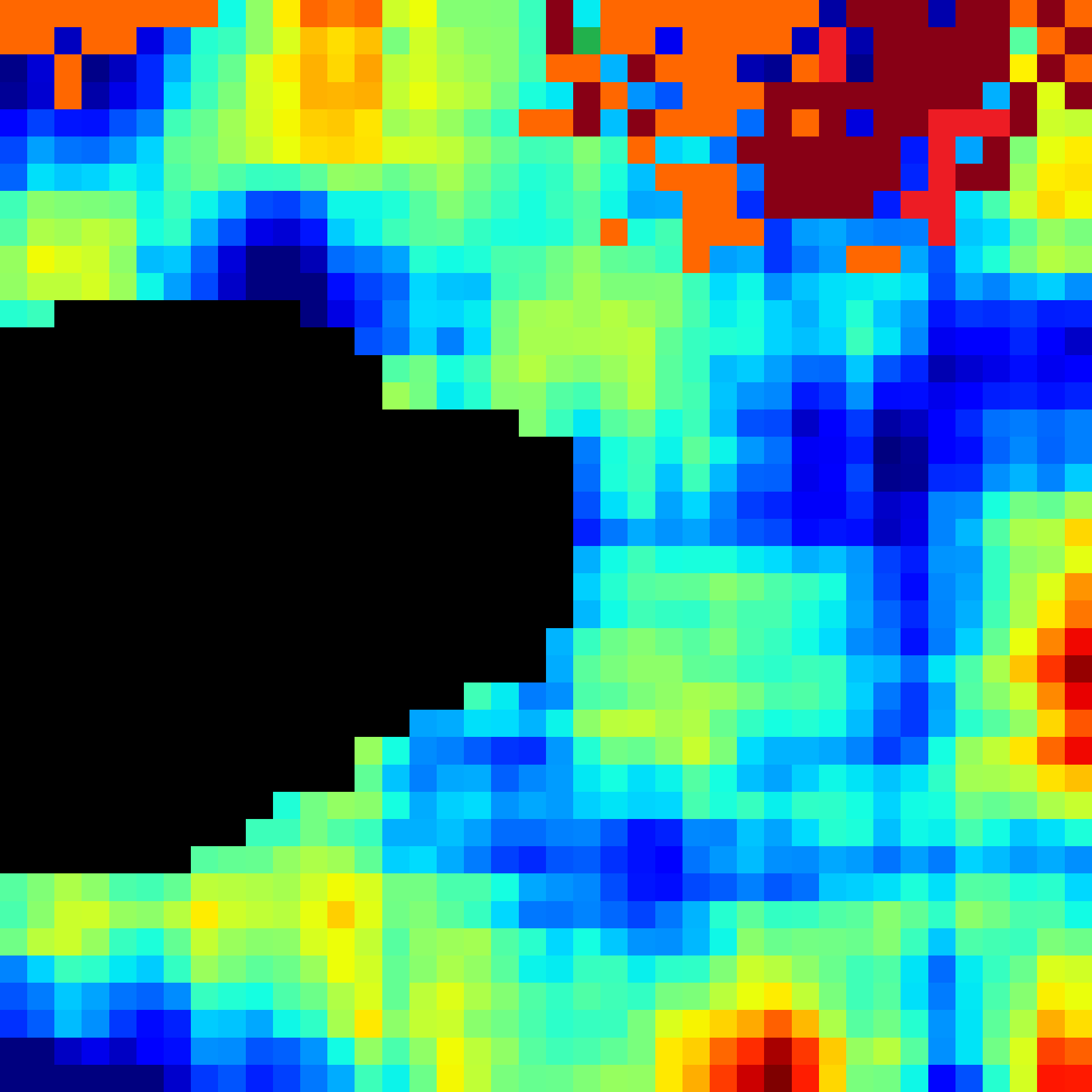}}
\caption{Estimated fish catch towards the coast of Easternmost point of Taiwan using (a) Sentinel-2 and (b) Sentinel-3 dataset. The blue to red colormap depicts the low to high estimated fish catch.}
\label{fig:predicted_catch}
\end{figure}

The estimated spatial map of fish catch is shown over the Easternmost point of Taiwan in Fig.~\ref{fig:predicted_catch}. The black color in both images depicts the masked landcover area. Surrounding this region, there is a strong spatial gradient of fish catch, transitioning from high-intensity catch near the North-East to low-intensity catch towards the South-West. The high catch around the North-East direction might be due to the oceanographic productivity differences. To the north, the oceanic current interacts with the continental shelf and submarine ridges, causing upwelling, eddies, and vertical mixing that bring cold, nutrient-rich deep water into the euphotic zone, boosting phytoplankton and zooplankton and thus attracting pelagic fish. In contrast, south of the easternmost point the oceanic current flows more smoothly and strongly offshore, forming a relatively oligotrophic condition.

\section{Conclusions}
\label{Sec:Conclusion}
This study finds that integrating an XGBoost-derived kernel into a Kernel Ridge Regression framework yields an accurate approach for satellite-based fish catch estimation using Sentinel-2 and Sentinel-3 data. Besides, spectral analysis indicates wavelength-dependent sensitivity to biological conditions at the ocean surface, with Sentinel-2 capturing fine-scale variability and Sentinel-3 providing insight into bulk water quality. Moreover, KRR-XGB outperforms linear and RBF kernels, achieving RMSE reductions of approximately 59–61\% for Sentinel-2 and 28–40\% for Sentinel-3, with strong correlations to in-situ observations ($\rho = 0.924$ and $0.731$). The resulting spatial fish catch maps exhibit physically consistent gradients aligned with known oceanographic processes, highlighting KRR-XGB’s potential for scalable, sensor-agnostic fishery monitoring.

\section*{Acknowledgement}
IIT Kharagpur supported the study through the project receipt grant No. IIT/SRIC/FSRG/2023/27


\bibliographystyle{IEEEtran}
\bibliography{mybibfile_n}

\end{document}